\documentclass{mn2e}

\usepackage{epsfig}
\usepackage{graphicx}
\usepackage{amssymb, latexsym, amsopn}

\title{Radiation in the process of the formation of voids}

\author[K. Bolejko]
{Krzysztof Bolejko$^{1}$
\thanks{E-mail: bolejko@camk.edu.pl}
\\
$^1$
Nicolaus Copernicus Astronomical Center, Polish Academy of Sciences,
ul. Bartycka 18, 00-716 Warsaw,
Poland
}

\pagerange{\pageref{firstpage}--\pageref{lastpage}}
\pubyear{2005}

\begin{document}

\maketitle
\label{firstpage}

\begin{abstract}
This paper aims to investigate the influence of inhomogeneous radiation in  void formation.
Since the process of void formation is non--linear, a fully relativistic model, which simulates the evolution of voids from the moment of the last scattering until the present instant, is presented. It is found  that in order to obtain a model  of a void which evolves from $\delta_0 = - 10^{-5}$
  at the last scattering moment to the present day $\delta = - 0.94$,
   the existence of radiation must be  taken into account.  
 The ratio of radiation energy density to matter energy density in ${\rm CDM}$ models at the moment of last scattering  is  $1/5$.
 This paper proves that
 such a  value of radiation energy density cannot be neglected and influences the first stages of void evolution. Namely, it is important to the process of structure formation and hence significantly influences the dynamics of the Universe
 in the first millions of years after  the last scattering.

From the fact that the evolution of voids proceeds differently in various cosmological background models we use the process of void formation to put some limits on values of cosmological parameters.
We find that the model with $\Omega_{mat} \sim 0.3$ fits the
 observational data best. 
\end{abstract}


\section{Introduction}

At the end of the 1970s
several galaxy redshift surveys started to operate. As a result the spatial distribution of a large sample of galaxies had been measured.
 It turned out that galaxies are distributed inhomogeneously and form structures like voids or clusters.

The most probable explanation of existence of such structures is that
they evolved from small initial fluctuations. These fluctuations can
be traced in the observations of the Cosmic Microwave Background
(CMB).

While gravity is an attractive force, it is relatively easy to reproduce high density regions, for instance by
setting the initial conditions so that collapse or shell crossings occur. This cannot be done in case of low density regions,  such as voids, inside which the estimated value of the density contrast is less than $\delta = -0.94$ (which is less than $6 \% $ of the mean background density) (Hoyle \& Vogeley 2004).
The description of  evolution of voids from initial fluctuations consistent with observational constraints is very difficult.
So far none of the attempts to solve this problem succeeded. Even N--body simulations predict that voids should be filled by dwarf galaxies which are not observed (Hoyle \& Vogeley 2004).

In this paper we are going to present a scenario of void formation which is consistent with astronomical observations and leads to the  formation of low density regions from very small initial fluctuations that existed at the moment of the last scattering. 

As will be demonstrated, the evolution of voids is a
non--linear process. That is why this approach is based on exact
solutions of the Einstein equations. 

The structure of this paper is as follows: in Sec. \ref{ssist} the model to be used  is  presented. In
Sec. \ref{coic} the initial conditions at the last scattering moment
are presented. These data constrain  values of parameters and functions 
used in the theoretical models. Since the influence of the gradient of  pressure cannot
be neglected, in Sec \ref{appox}  an  approximate and
an analytical solution of a model with inhomogeneous distribution of radiation is presented. This model qualitatively shows that the gradient of 
pressure plays a significant role in the evolution of the Universe
after the last scattering moment. In Sec.  \ref{result} 
 results of and fully non--linear evolution of models with inhomogeneous radiation distribution  are presented.  
 In Sec. \ref{cobm},  constraints on the values of cosmological parameters are derived from the fact 
that the structure formation proceeds differently in various cosmological
background models.

\section{The spherically symmetric inhomogeneous space--time}\label{ssist}

A spherically symmetric metric in comoving and synchronous
coordinates is of the form:

\begin{equation}
{\rm d}s^2 = {\rm e}^{A(t,r)} c^2 {\rm d}t^2 - {\rm e}^{B(t,r)}{\rm d}r^2
- R^2(t,r) \left({\rm d}\theta^2 + \sin^2 \theta {\rm d}\phi^2
\right). \label{ss}
\end{equation}

The Einstein field equations for the spherically symmetric perfect
fluid distribution (in coordinate components) are:

\begin{eqnarray}
G^{0}{}_{0} &=& {\rm e}^{-A} \left( \frac{\dot{R}^2}{R^2} + \frac{\dot{B} \dot{R}}{R} \right) - {\rm e}^{-B} \left( 2 \frac{R''}{R} \right. +  \nonumber \\
&+& \left. \frac{R'^2}{R^2} - \frac{B' R'}{R} \right) +
\frac{1}{R^2} = \kappa \epsilon + \Lambda,
\label{G00} \\
%%%%%%%%%%%%%%%%%%%%%%%%%%%%%%%%%%%%%%%
G^{1}{}_{0} &=& {\rm e}^{-B} \left( 2 \frac{\dot{R}'}{R} - \frac{\dot{B}
R'}{R} - \frac{A' \dot{R}}{R} \right) =0,
\label{G10} \\
%%%%%%%%%%%%%%%%%%%%%%%%%%%%%%%%%%%%%%%
G^{1}{}_{1} &=& {\rm e}^{-A} \left( 2 \frac{\ddot{R}}{R} + \frac{\dot{R}^2}{R^2} - \frac{\dot{A} \dot{R}}{R} \right) + \nonumber \\
&-& {\rm e}^{-B} \left(\frac{R'^2}{R^2} + \frac{A'
R'}{R}\right) + \frac{1}{R^2} = - \kappa p + \Lambda,
\label{G11} \\
%%%%%%%%%%%%%%%%%%%%%%%%%%%%%%%%%%%%%%%%%
G^{2}{}_{2} &=& G^{3}{}_{3} = \frac{1}{4} {\rm e}^{-A} \left( 4
\frac{\ddot{R}}{R}
 - 2 \frac{\dot{A} \dot{R}}{R} + 2 \frac{\dot{B} \dot{R}}{R} +  \right. \nonumber \\
 &+& \left. 2 \ddot{B} +  \dot{B}^2 - \dot{A} \dot{B} \right) - \frac{1}{4} {\rm e}^{-B} \left(4 \frac{R''}{R} + 2 \frac{A' R'}{R} + \right. \nonumber \\
 &-& \left. 2 \frac{B' R'}{R} + 2A'' + A'^2 - A' B' \right) = - \kappa p +
\Lambda, \label{G22}
\end{eqnarray}
where $\epsilon$ is the energy density and $p$ is the pressure. $\dot{}$ stands for $\partial_t$ and $'$ stands for $\partial_r$.  For
a mixture of matter and radiation, we have:

\begin{eqnarray}
\epsilon &=& \epsilon_{mat} + \epsilon_{rad}, \label{eps} \\
p &=& \frac{1}{3} \epsilon_{rad}.  \label{eos}
\end{eqnarray}

The Einstein equations can be reduced to (Lemaitre 1933):

\begin{equation}
 \kappa R^2 R' \epsilon = 2M',
\label{mr}
\end{equation}
\begin{equation}\label{emte}
 \kappa R^2 \dot{R} p = -2 \dot{M},
\label{mt}
\end{equation}
where $M$ is defined by:
\begin{eqnarray}\label{emdef}
 2M(r,t) &=& R(r,t) + R(r,t){\rm e}^{-A(t,r)}{\dot{R}}^2(r,t) + \nonumber \\
 &-& {\rm e}^{-B(t,r)}{R'}^2(r,t)R(r,t) - \frac{1}{3} \Lambda R^3(r,t).
\label{mdf}
\end{eqnarray}

In a Newtonian limit $M c^2/G$ is equal to the mass inside the shell of radial coordinate
$ r$.  However, it is not an integrated rest mass but  active gravitational mass that generates a gravitational field.
 As it can be seen in eq. (\ref{mdf}) the mass is not constant in time and in the expanding universe
it decreases, as seen from (\ref{mt}).

From the equations of motion ${T^{\alpha \beta}}_{; \beta} = 0$ we
obtain:
\begin{eqnarray}
 T^{0 \alpha}{}_{; \alpha} = 0 & \Rightarrow & \dot{B} + 4 \frac{\dot{R}}{R} = - \frac{2 \dot{\epsilon}}{ \epsilon + p},
\label{T0a} \\
 T^{1 \alpha}{}_{; \alpha} = 0 & \Rightarrow & A' = - \frac{2 p'}{ \epsilon + p},
\label{T1a} \\
T^{2 \alpha}{}_{; \alpha} = 0 & \Rightarrow &  \frac{\partial p}{\partial  \theta}, = 0, \label{praa} \\
T^{3 \alpha}{}_{; \alpha} = 0 & \Rightarrow & \frac{\partial p}{\partial  \phi} = 0.
\label{prad}
\end{eqnarray}

Equations (\ref{praa}) and (\ref{prad}) reproduce the well known
fact that the perfect fluid  energy--momentum tensor inherits the
symmetries of the metric of the space--time.

The $B$ function can be derived from eq. (\ref{G10}):

\begin{equation}
B(r,t) = B(r,t_0) + \ln R'^2 - \int\limits_{t_0}^t {\rm d \tilde{t}} 
\frac{A' \dot{R}}{R'}. \label{adrv}
\end{equation}

Using eq. (\ref{T1a}) we obtain:

\begin{eqnarray}
&& {\rm e}^{B(r,t)} = \frac{R'^2(r,t)}{1 + 2E(r)} \times \nonumber \\
&& \exp \left( \int\limits_{t_0}^t  {\rm d \tilde{t}}  \frac{2  \dot{R}(r, \tilde{t})}{ \left[ \epsilon(r,\tilde{t}) + p(r,\tilde{t}) \right] R'(r,\tilde{t})} p'(r,\tilde{t}) \right),
\label{edoA}
\end{eqnarray}
where $E(r)$ is an arbitrary function.

\subsection{The Lema\^{\i}tre--Tolman  model}

If the dust equation of state is considered, the above model becomes the
Lema\^{\i}tre-Tolman model. As follows from eq. (\ref{T1a}) and eq.
(\ref{edoA}), if  $p' = 0$ then:

\begin{eqnarray}
{\rm e}^{A} &=& 1, \\
{\rm e}^B &=& \frac{R'}{1+2E}.
\end{eqnarray}
 Then the  metric (\ref{ss}) becomes:

\begin{equation}
{\rm d}s^2 =  c^2{\rm d}t^2 - \frac{R'^2(r,t)}{1 + 2 E(r)}\ {\rm
d}r^2 - R^2(t,r) {\rm d} \Omega^2, \label{ds2}
\end{equation}
where $ {\rm d} \Omega^2 = {\rm d}\theta^2 + \sin^2 \theta {\rm
d}\phi^2$. Because of the signature $(+, -, -, -)$, the $E(r)$
function must obey $E(r) \ge - \frac{1}{2}.$

The Einstein  equations reduce to the following two:
\begin{equation}\label{den}
\kappa \rho c^2 = \frac{2M'}{R^2 R'},
\end{equation}
\begin{equation}\label{vel}
\frac{1}{c^2}\dot{R}^2 = 2E + \frac{2M(r)}{R} + \frac{1}{3} \Lambda
R^2,
\end{equation}
\noindent where $M(r)$ is another arbitrary function and $\kappa =
\frac{8 \pi G}{c^4}$.

When $R' = 0$ and $M' \ne 0$, the density becomes infinite. This
happens at shell crossings. This is an additional singularity to the
Big Bang that occurs at $R = 0, M' \neq 0$. Shell crossing can
be avoided by setting the initial conditions appropriately.

Equation (\ref{vel}) can be solved by simple integration:

\begin{equation}\label{evo}
\int\limits_0^R\frac{d\tilde{R}}{\sqrt{2E + \frac{2M}{\tilde{R}} +
\frac{1}{3}\Lambda \tilde{R}^2}} = c \left(t- t_B(r)\right),
\end{equation}
where $t_B$ appears as an integration constant, and is an arbitrary
function of $r$. This means that the Big Bang is not a single event
as in the Friedmann models, but occurs at different times at
different distances from the origin.

Thus, the evolution of a Lema\^{\i}tre-Tolman model is determined by
three arbitrary functions: $E(r)$, $M(r)$ and $t_B(r)$. The metric
and all the formulae are covariant under arbitrary coordinate
transformations of the form $r = f(r')$. Using such a
transformation, one function can be given a desired form. Therefore
the physical initial data for the evolution of the
Lema\^{\i}tre-Tolman model consist of two arbitrary functions.

\subsection{The Friedmann limit}\label{frdlmt}

Assuming  homogeneity, the above model becoms the Friedmann model.

 In the class of coordinates used here, one can choose the radial coordinate as:
\begin{eqnarray}
&& R(r,t) = r a(t), \label{r=a} \\
\end{eqnarray}
Then:
\begin{eqnarray}
&& M(r,t) = M_0 r^3, \\
&& 2 E(r) =  -k r^2. \\
&&{\rm e}^{A(r,t)} \to 1, \label{edac} \\
&&{\rm e}^{B(r,t)} \to \frac{a^2}{1-kr^2}, \label{edaa}  \\
&&\kappa \epsilon = \frac{6M}{R^3} \label{rdab}
\end{eqnarray}
where $a$ is the scale factor and $k$ is the curvature index of the
Friedmann models. Then the metric becomes:

\begin{equation}
{\rm d}s^2 = c^2 {\rm d}t^2 - a(t)^2 \left( \frac{{\rm d}r^2}{1 - k r^2} - r^2 \left({\rm d}\theta^2 + \sin^2 \theta {\rm d}\phi^2
\right) \right). \label{flrwds}
\end{equation}

The condition  $T^{\alpha \beta};_{\beta} = 0$, reduces to:

\begin{equation}
 \dot{\rho} + 3 \frac{\dot{a}}{a} \left( \rho + \frac{p}{c^2} \right)  = 0.
\label{1e}
\end{equation}

Assuming the dust equation of state,  eqs. (\ref{G11}), and (\ref{mdf})
 become the Friedmann equations:

\begin{eqnarray}
 && 2 \frac{ \ddot{a}}{a} = - \frac{1}{3} \kappa c^4 \rho  + \frac{2}{3} \Lambda c^2,  \label{fred} \\
&&  \dot{a}^2 = -k c^2 + \frac{1}{3} \kappa c^4 \rho a^2 + \frac{1}{3} \Lambda c^2 a^2  \label{fred2}
 \end{eqnarray}

The Friedmann limit is an essential element in our approach. As
mentioned above, our model of void formation describes a single void
in an expanding Universe. Far away from the origin, the density and
velocity distributions tend to the values that they would have in a
Friedmann model. Consequently the values of the time instants (i.e.
initial --- $t_0$ and final --- $t_f$ instants) and values of the
density  and velocity fluctuations are calculated with respect to
this homogeneous background.

 \section{Constraints on the initial conditions}\label{coic}

Voids are vast regions in which hardly any galaxies are observed. An
average radius of a void is $12~h^{-1}$ Mpc and the density contrast
inside is $-0.94$ (Hoyle \& Vogeley 2004). Such
structures must have evolved from small initial fluctuations
that existed at the last scattering instant. In this section the
estimation of the initial conditions is presented.

 \subsection{Observational constraints}\label{obsc}

 The observations of the CMB provide us with the  redshift of the surface of the last scattering.
Once the redshift is known,  density of  matter, density of radiation,  temperature and pressure can be estimated.  The measured value of the temperature fluctuations, $\Delta T \approx 10^{-5}$  can be converted into density and velocity fluctuations of a baryonic matter at that time.
 The estimated   amplitude of the initial fluctuations,
inside the voids, are (Bolejko, Krasi\'nski \& Hellaby 2005):

\begin{itemize}
\item
density fluctuations --- $ \Delta \rho / \rho = \delta = -6 \cdot
10^{-5} $
\item
velocity fluctuations --- $ \Delta V /V_b = \nu = 2 \cdot 10^{-4} $
\end{itemize}

The observed redshift of the CMB is $z \approx 1090$. At that
redshift  the amplitude of pressure of matter can be estimated from the
perfect gas equation of state:

\begin{equation}
\frac{p}{\rho c^2} = \frac{k_B T}{\mu m_H c^2} \approx 10^{-10}.
\end{equation}
This estimation shows that the pressure of  matter after the last scattering moment is negligible
in the evolution of the Universe.

The contribution of radiation  to the energy density cannot be neglected and  the ratio of radiation energy density to  matter energy density  at the last scattering moment is:

 \begin{equation}
 \frac{\epsilon_{rad}}{\epsilon_{mat}} = \frac{aT^4}{\rho c^2} = \frac{\Omega_{\gamma}}{\Omega_m} (1+ z),
 \label{rtest}
 \end{equation}
 where $a = 4 \frac{\sigma}{c}$ and $\sigma$ is Stefan--Boltzmann constant; $\Omega_{\gamma} = \frac{a T_{CMB}^4}{\rho_{cr}}$ and $T_{CMB}$ is the current temperature of the CMB background; $\rho_{cr}$ is critical density and is equal to $(3H_o) /(8 \pi G)$. 
 
At the moment of  last scattering, for $\Omega_{mat} = 0.3$ this is:
 \begin{equation}
 \frac{\epsilon_{rad}}{\epsilon_{mat}} \approx 0.2.
 \end{equation}
As can be seen, this ratio
decreases for low redshifts. However, at the early stage of the
evolution after the last scattering moment the radiation should have an
influence on the evolution of the Universe. This issue will
 be considered futher.

\subsection{Initial fluctuations in the linear approach}\label{laapp}

Although the present--day density contrast inside voids is large ($\delta \approx -0.9$) and it is unlikely that the linear approximation might handle it appropriately, the following calculations are presented to provide a comparison.
Let us assume that the Universe is  homogeneous with only  small
perturbations of the form:

\begin{eqnarray}
a &=& a_b (1 + \alpha(r,t))  \\
\rho &=& \rho_b (1 + \delta(r,t)).
\end{eqnarray}

Inserting these perturbations into eq. (\ref{fred})  and using eq. (\ref{1e})
 we obtain:

\begin{equation}
\ddot{\delta} + 2 \frac{\dot{a}_b}{a_b} \dot{\delta} - \frac{1}{2} \kappa c^2
\rho_b \delta = 0 
\label{linapp}
\end{equation}

In the simplest case, i.e. the Einstein--de Sitter model, this equation reduces to:
\begin{equation}
\ddot{\delta} + \frac{4}{3 c t} \dot{\delta} - \frac{2}{3 c^2 t^2} \delta = 0,
\end{equation}
and has the analytical solution:

\begin{equation}
 \delta = C_1 \tau^{2/3} + C_2 \tau^{-1},
 \label{linexct}
 \end{equation}
where $\tau$ is dimensionless time, i.e. $\tau = t/t_0$.
 
 As can be seen, the first factor describes the growing mode, while the second factor describes  the decaying mode. Assuming that only the growing mode  exists, we obtain:

\begin{equation}
 \delta = \delta_0 \tau^{2/3},
  \label{linsol}
  \end{equation}
where $\delta_0$ is the initial density contrast.

Assuming that the present-day density contrast of voids is $\delta =
- 0.9$, for the Einstein--de Sitter model,  $\tau = t_U/t_{LS} \approx 3.5
\times 10^{4}$ (where $t_U$ is the age of the Universe and $t_{LS}$ is the last scattering instant), we obtain:

\begin{equation}
  \delta_0 \approx 8 \times 10^{-4}.
   \end{equation}

However,   due to the  very 
large present-day value of the density contrast inside voids the linear approximation is inadequate.
Moreover,  the solutions of the formula (\ref{linapp})  have neither
upper, nor lower, limit. 
The mathematical structure of this equation allows the solution to be of any value, even of large negative value.
However from the physical point of view it is impossible to have a density contrast lower than $\delta_{lim} = -1$.
Unlike the linear approximation, the real density contrast inside the voids should 
asymptotically approach  $- 1$.
 Therefore, this relation is invalid in
investigation of void formation and an fully non--linear
solution must be considered.

\subsection{Initial fluctuations in the Lema\^itre--Tolman model}

Evolution of voids in the Lema\^itre--Tolman model was considered in
 Bolejko, Krasi\'nski and Hellaby
(2005). In that paper  the
evolution of a void from small initial velocity and density
fluctuations is investigated. The results imply that the existence of present--day
voids requires the amplitude of the initial density and velocity
fluctuations to be of $\delta \sim \nu \approx 5 \times
10^{-3}$.

\begin{figure}
\includegraphics[scale=0.7]{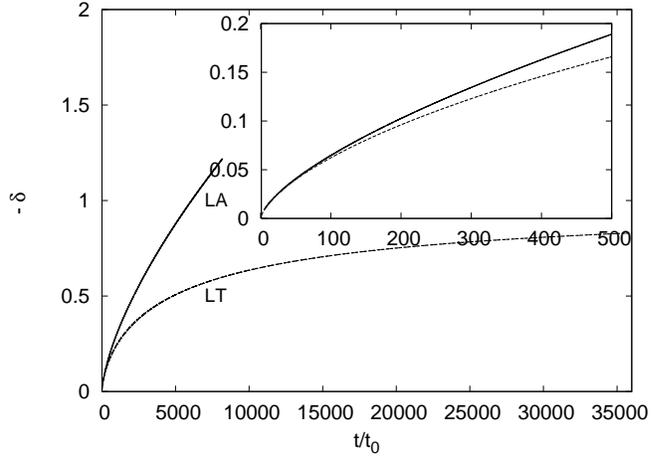}
\caption{The evolution of the density contrast in the linear approach and in the Lema\^itre--Tolman model. Both models are of the same initial fluctuations, $\delta_0 =  3 \times 10^{-3}$. For $\tau > 7000$ the linear approach predicts a density contrast lower than $-1$, which is not physical.}
\label{fig1}
 \end{figure}

 These results are by  one order of magnitude larger than the
 results obtained in the linear approach. Fig. \ref{fig1} presents the
 evolution of the density contrast in the linear approximation and in the
 Lema\^itre--Tolman model. Both models are of the same initial
 conditions, $\delta_0 = 3 \times 10^{-3}$. As  expected, the density contrast decreases more slowly in the non--linear model
 than in the linear approach. The difference between the results greater than $10 \%$ starts at $\tau \approx 500$, which corresponds to a redshift of $z \approx 15$. For redshifts larger than $z = 15$ the linear approach reproduces the results with error larger than $10 \%$, and for $\tau \approx 7000$ ($z \approx 2.25$) the density contrast is below $-1$ which is unphysical.

Such a disproportion between the theoretical prediction ($\delta_0 \sim 10^{-3}$) and
the observational measurements ($\delta_0 \sim 10^{-5}$) can be interpreted in two ways. 
The first alternative is  that the dark matter is made of some weakly interacting massive
particles, which at the moment of the last scattering could have
 the amplitude of  fluctuation of $10^{-3}$.
The second interpretation is that the matter fluctuations at the last scattering moment were of $10^{-5}$ and to obtain results consistent with observations some other effects must be taken into account. As was presented in Sec. \ref{obsc}, radiation is not negligible and might influence the evolution of voids. Although it seems rather unlikely that introduction of a small amount of radiation (temperature fluctuations at the last scattering moment were of $ \sim 10^{-5}$) could have a significant impact on void evolution, it will be shown below that in reality the radiation component plays an important dynamical role. It will be also shown that the linear approximation is inadequate here, and that the influence of radiation must be considered in the non--linear formalism of relativity.
In the next section an approximate model with inhomogeneous distribution of radiation is presented.

\section{Approximate solution}\label{appox}

As long as all fluctuations are very small the linear approach should be able to reproduce the proper results. Let us consider an approximate solution in the linear regime.

In comparison with Sec. \ref{laapp} let us consider the $\Omega_{mat} = 1,~\Omega_{\Lambda} = 0$ model with a small perturbation of radiation. Despite the  existence of radiation let us  assume that unperturbed quantities are the same as in the Einstein--de Sitter model:

 \begin{eqnarray}
&& k=0 \rightarrow E(r) = 0 \\
&& R_0 = r a_b = r a_0 \left(\frac{t}{t_0} \right)^{2/3} \\
&& \frac{\dot{a}_b}{a_b} = \frac{2}{3 t} \\
&& \rho = \rho_b = \widetilde{\rho_0} \left(\frac{t}{t_0} \right)^{-2} = \frac{1}{6 \pi G t^2}
\end{eqnarray}

 Futhermore, as only  the pressure fluctuations are of our interest let us assume that initial velocity and density fluctuations are equal to zero.
Radiation is as follows:

 \begin{eqnarray}
&& \xi = \xi_b (1 + \gamma) \\
&& \xi_b = \frac{1}{20} \widetilde{\rho_0} c^2 \left( \frac{t}{t_0} \right)^{-8/3} \label{xib} \\ 
&& p = \frac{1}{3} \xi = \frac{1}{3} \xi_b ( 1 + \gamma) = \frac{1}{3} \xi_b \left( 1 +  \gamma_0 \exp(- r^2/L^2) \right),
\end{eqnarray}
where $L^2$ is the size of the perturbed  region of radius  $5$ kpc at the last scattering moment.  $\gamma_0 = 8 \cdot 10^{-5}$ and is the amplitude of radiation fluctuation.
Based on eq. (\ref{rtest}) for $\Omega_{mat} = 1$ the  ratio of radiation energy density to matter energy density is $1/20$ and that is why in eq. (\ref{xib}) this value appears.

The density contrast is:

\begin{equation}
\delta = \frac{\rho - \rho_0}{\rho_0}
\end{equation}

In the Friedmann limit, eq. (\ref{rdab}):

 \begin{equation}
 \kappa c^2 \rho \rightarrow \frac{6 M}{R^3},~~~~~~~~~\kappa c^2 \rho_0 = \frac{6 M_0}{R_0^3} = 3 \left( \frac{\dot{R}_0}{R_0} \right)^2 =  \frac{\kappa c^2}{6 \pi G t^2}
\label{css}
\end{equation}

From eq. (\ref{mdf}) and the assumption that there are no density and velocity fluctuations:
\begin{equation}
\frac{2M}{R^3} = {\rm e}^{-A} \left(  \frac{\dot{R}_0}{R_0} \right)^2- {\rm e}^{-B(t,r)} \left( \frac{R_0'}{R_0} \right)^2  + \frac{1}{R_0^2} - \frac{1}{3} \Lambda
\end{equation}

In the following, we assume that $\Lambda = 0$ (at the last scattering and a few billion years after  the $\Lambda$ is negligible), and $A = 0$.

Under these assumptions:

\begin{equation}
\delta = - \frac{3 \frac{{\rm e}^{-B}{R'^2} - 1}{R^2}}{\frac{\kappa c^2}{6 \pi G t^2}}
\label{delss}
\end{equation}

Let us focus on the numerator first.
Using eq. (\ref{edoA}):

\begin{eqnarray}
 && \frac{{\rm e}^{-B(t,r)}{R'}^2(r,t) - 1}{R^2} = \frac{(1 + 2E(r))}{R^2} \times \nonumber \\
 && \exp \left( - \int\limits_{t_0}^t  {\rm d \tilde{t}} \frac{2  \dot{R}(r,\tilde{t})}{ \left[ \epsilon(r,\tilde{t}) + p(r,\tilde{t}) \right] R'(r,\tilde{t})} p'(r,\tilde{t}) \right) + \nonumber \\
 && -  \frac{1}{R^2} := \frac{ \mathcal{I} }{R^2}
\end{eqnarray}

Applying the assumption stated at the beginning of this section and assuming that:
\begin{equation}
\frac{p_0}{\rho_0 c^2 + (4/3) p_0} \approx \frac{p_0}{\rho_0 c^2},
\end{equation}

\begin{equation}
\mathcal{I} = \exp \left( - \int\limits_{t_0}^t  {\rm d \tilde{t}} \frac{2  \dot{R}(r,\tilde{t}) p'(r,\tilde{t})}{ \left[ \epsilon(r,\tilde{t}) + p(r,\tilde{t}) \right] R'(r,\tilde{t})}  \right) -1 
\end{equation}

Applying  series expansion to the $\exp$:

 \begin{equation}
 \mathcal{I} = \frac{1}{60} 2 \gamma_0 \frac{2r}{L^2} \exp(- \frac{r^2}{L^2}) \int\limits_{t_0}^t  {\rm d \tilde{t}} \frac{\rho_0 \tau ^{-8/3}}{ \left[ \rho_0 \tau^{-2} \right]} \frac{\dot{R}}{R'}
 \end{equation}

 \begin{equation}
 \mathcal{I} = \frac{1}{60} 2 r \gamma_0 \frac{2r}{L^2} \exp(- \frac{r^2}{L^2}) \int\limits_{1}^{\tau}  {\rm d \tau'} \frac{\rho_0 \tau ^{-8/3}}{ \left[ \rho_0 \tau^{-2} \right]} \frac{2}{3 \tau}
 \end{equation}

Close to the origin, $r \approx  0$:

 \begin{equation}
 \mathcal{I} = \frac{8}{180}  r^2 \frac{\gamma_0}{L^2}  \int\limits_{1}^{\tau}  {\rm d \tau'}  \tau ^{-5/3}
 \end{equation}

 \begin{equation}
 \mathcal{I} = \frac{1}{15}  r^2 \frac{\gamma_0}{L^2}  (1 -  \tau ^{-2/3})
 \end{equation}

As can be seen,  $ \mathcal{I}$ has the maximal value, :

\begin{equation}
  I_{max} = \frac{1}{15}  r^2 \frac{\gamma_0}{L^2}
 \end{equation}

 Inserting the above to eq. (\ref{delss}):

 \begin{equation}
 \delta = - 3 \frac{\frac{\mathcal{I}}{R^2}}{\frac{\kappa c^2}{6 \pi G t^2}} = - \frac{3}{20}    \frac{c^2 t_0^2}{L^2 a_0^2} \tau^{2/3}
 \end{equation}

 To calculate the  value of the coefficient, we assume that at $t_0 = t_{LS}$,  $a_0 = 1$. Then:

 \begin{equation}
 \delta = - 3 \times 10^{-3} \tau^{2/3}
 \end{equation}

 Comparing this result with eq. (\ref{linsol}) it is easy to see that
 introducing only pure radiation fluctuation, the results are the same as the results obtained in the linear approach without any radiation component. The corresponding initial density fluctuation in the linear approach would be:

 \begin{equation}
 \delta_0 = - 3 \times 10^{-3}
 \end{equation}

This proves that radiation is of great importance in
voids evolution. However  to be able to obtain  precise
predictions, more accurate calculations are needed. The evolution of voids within an non--linear model is presented in next section.

\section{The algorithm}

The computer algorithm used to calculate the evolution of a void was written in
Fortran and consisted of the following steps.  Numerical methods are from Press
et  al. (1986) and Pang (1997).

\subsection{The initial data}

\subsubsection{Time instants}

As follows from Sec. \ref{ssist} to specify the model, one
needs to know three functions.
In this case these function are:
 density, velocity and radiation distribution functions, 
and are presented in subsections below.
Because it is required that at large distace from the origin
the model of a void becomes 
an homogeneous Friedman model, these
 functions are presented in form of 
given fluctuations imposed on the homogeneous background.
All background values are calculated for the time instants obtained from the 
follwing formula (Peebles 1993):

\begin{equation}
t(z) =  \frac{1}{H_o} \int\limits_{z}^{\infty} \frac{d \tilde{z}}{(1+\tilde{z}) \sqrt{ \mathcal{D}(\tilde{z})} },
\label{tz}
\end{equation}
where:
\begin{equation}
\mathcal{D} (z) = \Omega_{\gamma}(1+z)^4 + \Omega_{mat} (1+z)^3 + \Omega_K (1+z)^2 + \Omega_{\Lambda},
\end{equation}
 $H_o$ is the present Hubble constant, $\Omega_K = 1 - \Omega_{\gamma} - \Omega_{mat} - \Omega_{\Lambda}$.
The initial instant ($t_0$) is set to be at  the last scattering  moment, which
 took place when $z \approx 1089$ (Bennett et al. 2003) and the final instant ($t_f$) when $z=0$.

\subsubsection{The initial density perturbations}

The radial coordinate was defined as the areal radius at the moment
of last scattering, measured in kiloparsecs:

 \[   r = \ell = R_{0}/d = R(r, t_{0})/d   \]
where $d = 1$~kpc.

The initial density  fluctuations, imposed on the
homogeneous background, were defined by functions
of the radius $\ell$,
 as listed in Tables \ref{InPer} and \ref{InPer2}, and the actual
density fluctuations followed from:

\begin{equation}
 \rho_0(\ell) = \rho_{b{}0} \cdot \delta(\ell),
\end{equation}
where $\rho_{b{}0}$ is the density of the homogeneous  background and at the initial instant, and can be expressed as:

\begin{equation}
\rho_{b{}0} = \Omega_{mat} \rho_{cr} (1+z)^3.
\end{equation}

The  mass  inside the shell of radius $R(r,t_0)$ at the initial instant, measured in
kiloparsecs, was calculated by integrating eq. (\ref{mr}):
 \begin{equation}
   M(\ell, t_0) - M(0, t_0) = \left. \frac{\kappa c^2}{2} \int\limits_{\ell_{min}}^{\ell}
   \rho_i(\ell') \ell'^2 d\ell' \right|_{t = t_{1}} . \label{ModR}
 \end{equation}
Since the density distribution has no singularities or zeros over extended
regions, it was assumed that $\ell_{min} = 0$ and $M = 0$ at $\ell = 0$.

\subsubsection{The initial velocity perturbations}

The initial velocity  fluctuations, imposed on the
homogeneous background, were defined by functions
of the radius $\ell$,
 as listed in Tables \ref{InPer} and \ref{InPer2}, and the actual
velocity fluctuations  followed from:

\begin{equation}
U(\ell) = U_{b{}0} \cdot \nu(\ell),
\end{equation}
where $U_{b{}0}$ is the velocity of
the homogeneous  background at the initial instant, and can be expressed, from eq. (\ref{r=a}),  as:

\begin{equation}
U_b = \frac{1}{c} \dot{R} = \frac{1}{c} r \dot{a}.
\end{equation}

In the FLRW models the time derivative of the scale factor is given by the formula (Peebles 1993):

\begin{equation}
\dot{a} = a H_o \sqrt{\mathcal{D}}.
\end{equation}

Consequently:

\begin{equation}
U = (1 + \nu) \frac{1}{c} R H_o \sqrt{\mathcal{D}}.
\label{exs}
\end{equation}

In the Lema\^itre--Tolman models the proper--time derivative of the areal radius $R$ ($U$) is just equal to $ \frac{1}{c} \dot{R}$, in our case, in consequence of  the metric (\ref{ss}):
\begin{equation}
U = \frac{1}{c} {\rm e}^{-A/2}  \dot{R}.
\label{udf}
\end{equation}

\subsubsection{The radiation perturbations}

In FLRW models the equations of motion reduce to (Pleba\'nski \& Krasi\'nski 2005):

\begin{equation}
\partial_t (\epsilon_{} a^3) + 3 a^2 \dot{a} p = 0
\end{equation}

For matter and radiation background (eqs.
(\ref{eps}) and (\ref{eos})) this equation becomes (to simplify the notation, we denote $\epsilon_{rad} = \mathcal{E}$ and $\epsilon_{mat} = \rho$):

\begin{equation}
( \dot{\mathcal{E}} + \dot{\rho}) a^3 + 3 (\mathcal{E} + \rho) a^2 \dot{a} + a^2 \dot{a} \mathcal{E} = 0.
\label{feom}
\end{equation}

After the last scattering moment the radiation has not been interacting with matter,
so we can assume that the evolution of matter is the same as in the Universe without radiation, i.e.:
\[ \rho = \rho_o \left( \frac{a_o}{a} \right)^3. \]
Then from (\ref{feom}):
\[ \mathcal{E} = \mathcal{E}_o \left( \frac{a_o}{a} \right)^4. \]

 In the inhomogeneous Universe the radiation can be written in the following form:

\begin{equation}
\epsilon_{rad}(r,t) = \mathcal{E}(t_f) (1+z)^4 \zeta(r,t),
\end{equation}
where $\zeta(r,t)$ is the function which describes the distribution and the evolution of radiation.

According to the current paradigm,  after the last scattering moment  the distribution of
radiation has been "frozen".
Consequently the only change in the radiation distribution is in the amplitude, which decreases with time, as the Universe expands:

\begin{equation}
\zeta(r,t) = \phi(t) \gamma(r)
 \label{rA1}
 \end{equation}

The crucial problem is to find the form of the $\phi$ function.
Luckily, the fluctuations of the radiation are very small, of amplitude $\sim 10^{-5}$, therefore, we can assume that the time--dependent amplitude of the radiation is the same as in the homogeneous background, so:

\begin{equation}
\phi = 1.
\label{rA2}
\end{equation}

However, this assumption may have to be modified in the future if observational data on the distribution of radiation become more detailed.

Recapitulating:

\begin{equation}
\epsilon_{rad}(\ell, z(t)) = 4 \frac{\sigma}{c} T^4_{CMB} (1+z)^4 \cdot \gamma(\ell)),
\label{rb1}
\end{equation}
where $\sigma$ is the Stefan-Boltzmann constant, $c$ is the speed of light, $T_{CMB}$ is the current measured temperature of the CMB and
$\gamma$ describes the initial perturbation of radiation, as listed in Tables \ref{InPer} and \ref{InPer2}.

Solving (\ref{tz}) with the bisection method the value of $z$ for some instant $t$
can be found.
Substituting this value to (\ref{rb1})  the value of energy density of radiation for this instant can be calculated.

\subsection{Computing the evolution}

The algorithm of the evolution consisted of the following steps:

\begin{enumerate}
\item
From eqs. (\ref{udf}) and (\ref{T1a}) the value of $\dot{R}$  and from (\ref{mt}) the value of $\dot{M}$ can be calculated.
Then using the predictor--corrector method the value of $R(r,t+ \tau)$ and $M(r, t+ \tau) $ in the time step of $\tau$ can be found. We futher denoted them (as all the quantities found in this time step) by the subscript $_\tau$.
\item
Once $R_\tau$ and $M_\tau$ are known, we can derive $\epsilon_\tau$ from (\ref{mr}).
\item
Then from (\ref{eos}) and (\ref{rb1}) we derive $p_\tau$.
\item
 From $\epsilon_\tau$ and $p_\tau$ we can calculate $A_\tau$ by integrating eq. (\ref{T1a}).
\item

$U_\tau$ can be calculated as follows:

From eq. (\ref{adrv}) and (\ref{mdf}) we obtain:

\begin{eqnarray}
&& U^2(r,t)  = \frac{2M(r,t)}{R(r,t)} + \frac{1}{3} \Lambda R^2(r,t) -1  \nonumber \\
&&  + {\rm e}^{-B(r,t_0)}{\rm e}^{\int_{t_0}^{t}  {\rm d \tilde{t}} \left( A'(r,\tilde{t}) {\rm e}^{A(r,\tilde{t})/2} U(r,\tilde{t}) \right) /R'(r,\tilde{t})}.
  \end{eqnarray}

By solving this equation by the bisection method, for the time $t=t_0+\tau$ we can calculate $U_\tau$.

\item
Once $U_\tau$ and $A_\tau$ are known, $\dot{R}(r,t)$ and futher $\dot{M}(r,t)$ can be calculated.
\item
We repeat steps 1--6 until $t=t_f$.

\end{enumerate}

\section{Results}\label{result}

The  measurements of  density contrast inside voids 
are based on the observations of galaxies inside them (Hoyle \& Vogeley 2004).
However, because in central regions no galaxies are observed, the real density distribution inside the voids is unknown. Assuming that luminous matter is a good tracer of dark matter distribution and extrapolating the value of the density contrast measured on the edges of voids (where galaxies are observed) into the central regions of voids, we can conclude that
the density inside the voids is below the value of $0.06 \rho_b$. This value will be called the limiting value. It is expected that the model will predict a present--day density inside the voids below this limiting value.

While the problem of choice of the cosmological model is open and the efforts to   determine the values of $\Omega_{mat}$ and $\Omega_{\Lambda}$ still continue, in this section (except subsection \ref{cobm}) we focused only on the cosmological model which is the most popular in the cosmological comunity, that is $\Lambda {\rm CDM}$ model.

\begin{figure}
\includegraphics[scale=0.7]{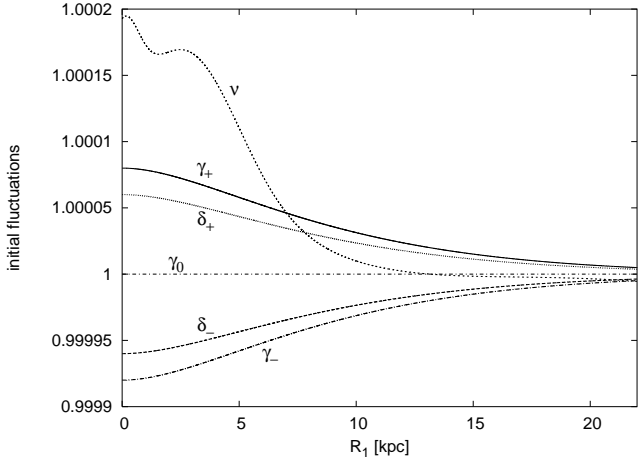}
\caption{The shape of the initial fluctuations.}
\label{init}
 \end{figure}

Figure \ref{init} shows the shape of the initial perturbations. The explicit forms of these perturbations are presented in Tables \ref{InPer} and \ref{InPer2}.
The results are presented in Fig \ref{fin}. In four out of seven model voids were formed.
As can be seen, models with inhomogeneous distribution of radiation have no problems
with reproducing regions of density below the limiting value.

To compare our results with the observational data, Fig. \ref{obs} presents the average density contrast inside the voids as a function of a relative  distance from the origin. The average density contrast was calculated as follows:

\begin{equation}
 \delta = \frac{ < \rho> }{\overline{\rho}} - 1,
 \label{dcon}
 \end{equation}
 where $\overline{\rho}$ is the present background density and $<\rho>$:

 \begin{equation}
 < \rho> = \frac{ 3 M c^2}{ 4 \pi G R^3}.
 \end{equation}

Curve {\bf 1} presents the results of run 1, as listed in Table \ref{InPer}.
Curves \textsc{\textbf{NGP}} and \textsc{\textbf{SGP}} corespond to
 density contrasts of voids in the 2dFGRS data estimated by Hoyle \& Vogeley.
Although the profiles match at the center they do not fit accurately at the edges of voids.
In our model the density contrast tends to increase faster than the observed one which
could be caused by too strong assumptions about the evolution of  radiation ((\ref{rA1}) and
(\ref{rA2})) and would suggest that the distribution of radiation did evolve after the last scattering moment.
There is another possibility that could explain the difference between these two profiles.
The density contrast  estmated by Hoyle \& Vogeley is based on observations of galaxies.
It is possible that the existence of  dark matter in walls would make the real density higher than that estimated from counting of galaxies.

\subsection{Initial perturbations}

Introducing radiation into the calculation we need to know the relation between matter and radiation perturbations.
In linear theory there are three concepts of these relations:

\begin{enumerate}
\item
adiabatic perturbations, where $\gamma = \frac{4}{3} \delta$,
\item
isocurvature perturbations, where $\gamma = \frac{4}{3} ( \delta - \delta_o)$; ($\delta_o$ is some initial value of $\delta$)
\item
isothermal perturbation, where $\gamma = 0$.
\end{enumerate}

It should be stressed that in realistic conditions there are no pure adiabatic or isocurvature, or isothermal perturbations and the relations between density and radiation perturbations are more complicated.
However, it is instructive to know
  what kind of relation is more suitable for the process of void formation.

\begin{table}
\caption{ \label{InPer} The initial Density Perturbations used in the
runs.    }
\begin{tabular}{lll}
Run & Profile  &  Description   \\ \hline
%%%%%%%%%%%%%%%%%%%%%%%%%%%%%%%%%%%%%%%%%%%%%%%%%%%%%%%%%%%%%%%%%%%%%%%
1 &  $ \delta_- = 1 - \mathcal{F}  $    &   Isocurvature--like perturbation.    \\
  &  $\gamma_+ = 1 + \frac{4}{3} \mathcal{F} $  &   Reconstructed the present-day voids.    \\
  & $ \nu = 1 + \mathcal{G}  $ &         \\ \hline
%%%%%%%%%%%%%%%%%%%%%%%%%%%%%%%%%%%%%%%%%%%%%%%%%%%%%%%%%%%%%%%
2 &  $ \delta_- = 1 - \mathcal{F}  $    &   Adiabatic  perturbation.    \\
  &  $\gamma_- = 1 - \frac{4}{3} \mathcal{F} $  &   The collapse after 20 millions years.   \\
  & $ \nu = 1 + \mathcal{G}  $ &   Leads to  high--density regions.              \\ \hline
%%%%%%%%%%%%%%%%%%%%%%%%%%%%%%%%%%%%%%%%%%%%%%%%%%%%%%%%%%%%%%%
3 &  $ \delta_- = 1 - \mathcal{F}  $    &   Isothermal perturbations.   \\
  &  $\gamma_0 = 1  $   &   Do not lead to low--density region.    \\
  & $ \nu = 1 + \mathcal{G}  $ &                 \\ \hline
%%%%%%%%%%%%%%%%%%%%%%%%%%%%%%%%%%%%%%%%%%%%%%%%%%%%%%%%%%%%%%%
4 &  $ \delta_+ = 1 + \mathcal{F}  $    &   Adiabatic  perturbation.    \\
  &  $\gamma_+ = 1 + \frac{4}{3} \mathcal{F} $  &   Reconstructed the present-day voids.    \\
  & $ \nu = 1 + \mathcal{G}  $ &                 \\ \hline
%%%%%%%%%%%%%%%%%%%%%%%%%%%%%%%%%%%%%%%%%%%%%%%%%%%%%%%%%%%%%%%
5 &  $ \delta_+ = 1 + \mathcal{F}  $    &  Reconstructed the present-day voids, \\
  &  $\gamma_+ = 1 + \frac{4}{3} \mathcal{F} $  & although the density fluctuations are  \\
  & $ \nu_- = 1 - \mathcal{G}  $ & positive and velocity perturb. are negative.                  \\ \hline
%%%%%%%%%%%%%%%%%%%%%%%%%%%%%%%%%%%%%%%%%%%%%%%%%%%%%%%%%%%%%%%
6 &  $ \delta_0 = 1  $  &  Reconstructed the present-day voids. \\
  &  $\gamma_+ = 1 + \frac{4}{3} \mathcal{F} $  &  \\
  & $ \nu_0 = 1  $ &                   \\ \hline
%%%%%%%%%%%%%%%%%%%%%%%%%%%%%%%%%%%%%%%%%%%%%%%%%%%%%%%%%%%%%%%
7 &  $ \delta_- = 1 - \mathcal{F}  $    &  Leads to the cluster formation, \\
  &  $\gamma = \delta_- = 1 -  \mathcal{F} $    & with the value of the central density  \\
  & $ \nu = 1 + \mathcal{G}  $ &  $21 \cdot 10^{3}  \rho_b$.    \\ \hline
%%%%%%%%%%%%%%%%%%%%%%%%%%%%%%%%%%%%%%%%%%%%%%%%%%%%%%%%%%%%%%%

\end{tabular}
\end{table}

\begin{table}
\caption{ \label{InPer2} Detailed descriptions of the initial fluctuations from Table \ref{InPer}. All the values in the table are dimensionless, and the distance parameter
is the areal radius in kiloparsecs $\ell = R_i$/1kpc.
  }
\begin{tabular}{ll}
  Profile  & Parameters    \\ \hline
%%%%%%%%%%%%%%%%%%%%%%%%%%%%%%%%%%%%%%%%%%%%%%%%%%%%%%%%%%%%%%%%%%%%%%%
  $ \mathcal{F}(\ell) = a \cdot f^c \cdot {\rm e}^{d \cdot f} + g$ & $a = (A-g) \cdot (-b)^{-c} \cdot {\rm e}^c $       \\
%%%%%%%%%%%%%%%%%%%%%%%%%%%%%%%%%%%%%%%%%%%%%%%%%%%%%%%%5
 & $f = \ell - b $          \\
%%%%%%%%%%%%%%%%%%%%%%%%%%%%%%%%%%%%%%%%%%%%%%%%%%%%%%%%5
 &  $ g = A \cdot \left[ 1 - (-b)^c \cdot {\rm e}^{-c} \cdot h \right]^{-1} $                        \\
 %%%%%%%%%%%%%%%%%%%%%%%%%%%%%%%%%%%%%%%%%%%%%%%%%%%%%%%%5
&  $ h = (R - b)^c {\rm e}^{d(b-R)} $    \\
%%%%%%%%%%%%%%%%%%%%%%%%%%%%%%%%%%%%%%%%%%%%%%%%%%%%%%%%5
 & $ c = d \cdot b  $    \\
%%%%%%%%%%%%%%%%%%%%%%%%%%%%%%%%%%%%%%%%%%%%%%%%%%%%%%%%5
& $ d=-0.2 $     \\
%%%%%%%%%%%%%%%%%%%%%%%%%%%%%%%%%%%%%%%%%%%%%%%%%%%%%%%%5
&  $ b = -4.6 $  \\
%%%%%%%%%%%%%%%%%%%%%%%%%%%%%%%%%%%%%%%%%%%%%%%%%%%%%%%%5
 & $ R =40 $     \\
 %%%%%%%%%%%%%%%%%%%%%%%%%%%%%%%%%%%%%%%%%%%%%%%%%%%%%%%%5
& $ A=6 \cdot 10^{-5} $  \\
%%%%%%%%%%%%%%%%%%%%%%%%%%%%%%%%%%%%%%%%%%%%%%%%%%%%%%%%5
%%%%%%%%%%%%%%%%%%%%%%%%%%%%%%%%%%%%%%%%%%%%%%%%%%%%%%%%5
    \hline
  $  \mathcal{G}(\ell) = A \cdot ( b \cdot \arctan{c} - d \cdot
 \ell - $ & $  A = - 1 \cdot 10^{-4} $  \\
 %%%%%%%%%%%%%%%%%%%%%%%%%%%%%%%%%%%%%%%%%%%%%%%%%%%%%%%%5
    $~~~~~~~~   f \cdot e^{-g^2}  - e^{-h^2}  - e^{-j^2} - $  &  $ b = 4  $   \\
  %%%%%%%%%%%%%%%%%%%%%%%%%%%%%%%%%%%%%%%%%%%%%%%%%%%%%%%%5
    $~~~~~~~~ m \cdot e^{-n^2}) \cdot k + p $ &  $ c = 0.02 \cdot \ell - 0.02 $ \\
  %%%%%%%%%%%%%%%%%%%%%%%%%%%%%%%%%%%%%%%%%%%%%%%%%%%%%%%5
     &  $ d = \frac{5}{55} $  \\
  %%%%%%%%%%%%%%%%%%%%%%%%%%%%%%%%%%%%%%%%%%%%%%%%%%%%%%%%5
     & $ f= 0.7 $  \\
  %%%%%%%%%%%%%%%%%%%%%%%%%%%%%%%%%%%%%%%%%%%%%%%%%%%%%%%%5
     & $ g=  \ell $  \\
  %%%%%%%%%%%%%%%%%%%%%%%%%%%%%%%%%%%%%%%%%%%%%%%%%%%%%%%5
     & $ h = \frac{\ell-1}{7} $  \\
  %%%%%%%%%%%%%%%%%%%%%%%%%%%%%%%%%%%%%%%%%%%%%%%%%%%%%%%5
     & $ j = \frac{\ell-3}{3} $  \\
  %%%%%%%%%%%%%%%%%%%%%%%%%%%%%%%%%%%%%%%%%%%%%%%%%%%%%%%5
     & $ m= 1.225 $  \\
  %%%%%%%%%%%%%%%%%%%%%%%%%%%%%%%%%%%%%%%%%%%%%%%%%%%%%%%5
     & $ n= \frac{\ell - 39}{12} $  \\
  %%%%%%%%%%%%%%%%%%%%%%%%%%%%%%%%%%%%%%%%%%%%%%%%%%%%%%%5
     & $ k= \frac{1}{1 + 0.03 \ell} $  \\
  %%%%%%%%%%%%%%%%%%%%%%%%%%%%%%%%%%%%%%%%%%%%%%%%%%%%%%%5
     & $ p = - 2 \times 10^{-5} $  \\
  %%%%%%%%%%%%%%%%%%%%%%%%%%%%%%%%%%%%%%%%%%%%%%%%%%%%%%%%5
    \hline
\end{tabular}
\end{table}

\begin{figure}
\includegraphics[scale=0.7]{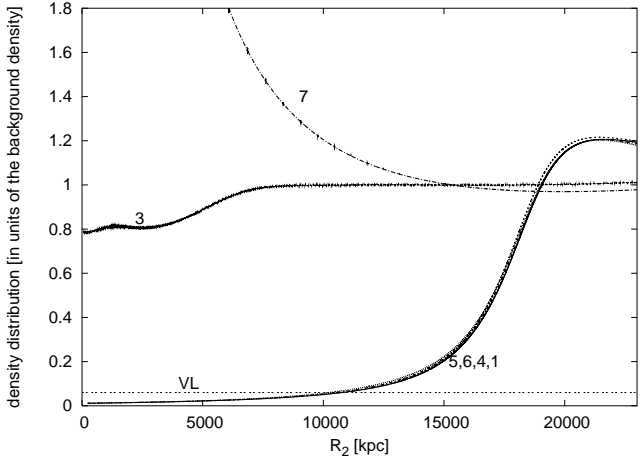}
\caption{The density distribution inside the void. The numbers of curves correspond to the numbers of runs (description in Table \ref{InPer}) in the $\Lambda {\rm CDM}$ background model. The curve denoted by {\bf \textsc{vl}} (void limit) refers to the measured density inside the voids.}
\label{fin}
 \end{figure}

\begin{figure}
\includegraphics[scale=0.7]{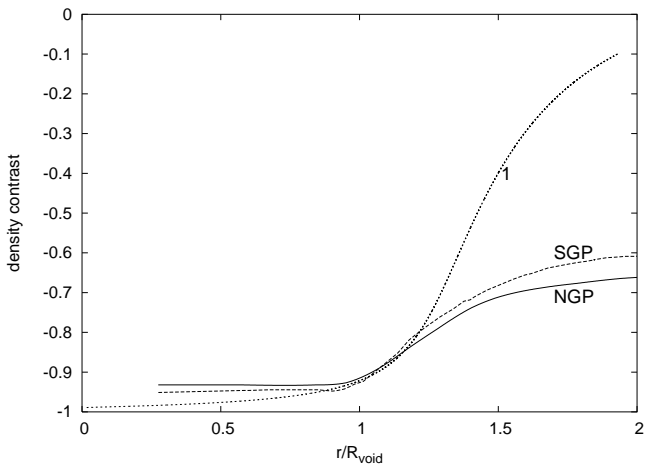}
\caption{Comparison of the  density contrast obtained in run 1 ({\bf 1}) with
the observed density contrast from Hoyle and Vogeley (2004) ({\bf \textsc{SGP}}) and
({\bf \textsc{NGP}}).}
\label{obs}
\end{figure}

The results presented in Figure \ref{fin}
imply that voids can be formed out of adiabatic or isocurvature perturbations
and there is no significant difference between these two forms of perturbations, as long as
the gradient of the radiation is negative.
 With an isothermal perturbation
low density regions cannot be formed as the gradient of radiation is important in the process of void formation.

\subsection{Validity of the linear approach}\label{val}

\begin{figure}
\includegraphics[scale=0.7]{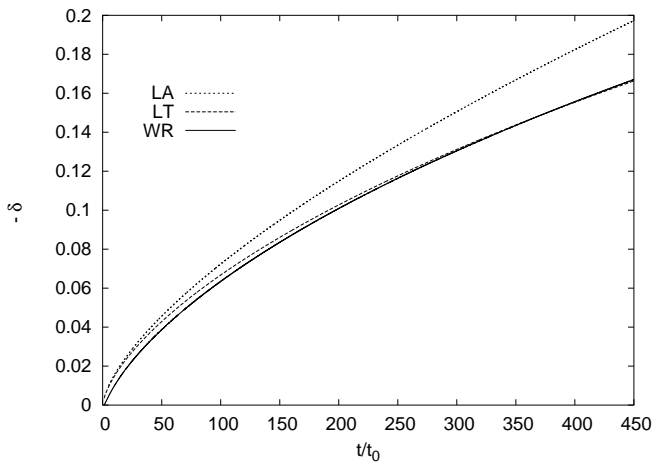}
\caption{Comparison of the  density contrast evolution in the linear regime (upper line), Lema\^itre--Tolman model (middle line) and the model with only radiation fluctuations. The initial fluctuation in the linear regime and in LT model are $\delta_0 = 3 \times 10^{-3}$. In the model with only radiation fluctuation, the initial radiation perturbations are $\gamma_0 = 8 \times 10^{-5}$.}
\label{dce}
\end{figure}

Fig. \ref{dce} presents the evolution of the density contrast in three models: in the linear approximation, in the Lema\^itre--Tolman model, and in an inhomogeneous model with  radiation perturbation only. The first two models are of the same initial condition. As one can see, the linear approach, given by (\ref{linapp}), is inadequate. At the early stages   radiation drives the evolution. Because this effect is not considered in the linear approach, the results obtained in the linear approach are not accurate.  For later times, when the radiation becomes negligible, the density contrast is  too large to be correctly handled by the linear approximation.

\subsection{Constraints on the background model}\label{cobm}

\begin{figure}
\includegraphics[scale=0.7]{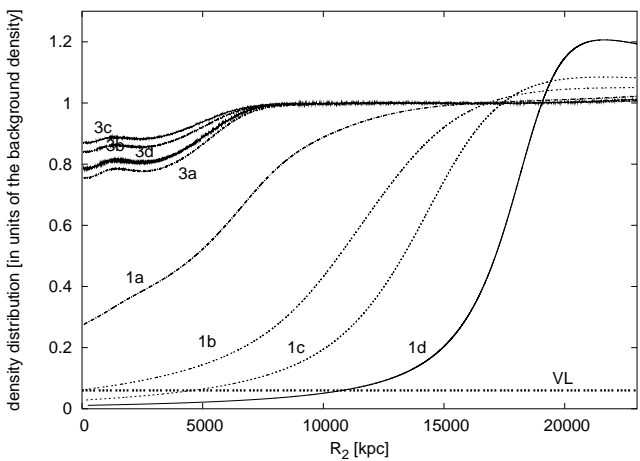}
\caption{The density distribution inside the void with three different background models. The numbers of curves corespond to the numbers of runs (description in Table \ref{InPer}) and to the background model (as listed in sec. \ref{cobm}). The curve denoted by {\bf \textsc{VL}} (void limit) refers to the measured density inside  voids.}
\label{bgmod}
 \end{figure}

Figure \ref{bgmod} presents the evolution of  voids
(the initial  conditions like in  runs 1 and 3)
 in four different background models:

\vspace{0.25cm}

(a)~~   $\Omega_{mat} = 1$, $\Omega_{\Lambda} = 0$, $\Omega_{\gamma} = 4.77 \cdot 10^{-5}$,
\vspace{0.1cm}

(b)~~  $\Omega_{mat} = 0.4$, $\Omega_{\Lambda} = 0$, $\Omega_{\gamma} = 4.77 \cdot 10^{-5}$,

\vspace{0.1cm}

(c)~~  $\Omega_{mat} = 0.27$, $\Omega_{\Lambda} = 0$, $\Omega_{\gamma} = 4.77 \cdot 10^{-5}$,

\vspace{0.1cm}

(d)~~  $\Omega_{mat} = 0.27$, $\Omega_{\Lambda} = 0.73$, $\Omega_{\gamma} = 4.77 \cdot 10^{-5}$.

\vspace{0.25cm}

These results imply that in the absence of radiation, or of the gradient of radiation, the structure formation goes on faster in the models which are filled with a greater amount of  matter (curves {\bf 3a}, {\bf 3b}, {\bf 3c} and {\bf 3d} --- for more details see Bolejko, Krasi\'nski \& Hellaby 2005).
Voids cannot be formed within this kind of  radiation perturbations. By introducing a realistic distribution of  radiation
 voids are formed more likely.

According to the astronomical observations, about $40 \%$ of the volume of the Universe is taken up by voids (Hoyle and Vogeley
2004). This means that void formation is
not an isolated event, but it is a very probable process. Thus, it can
 be used to put some constraints on the cosmological background model.

The results in Fig. \ref{fin} imply that the presence of  radiation is important for void formation.
 The contribution from radiation to the evolution of the system is more significant in the models with smaller value of $\Omega_{mat}$.
Therefore, the process of void formation constrains the value of $\Omega_{mat}$.

As can be seen in Fig. \ref{bgmod}, the cosmological constant is not needed
to reconstruct the present--day voids, thus  the process of void formation does not constrain the cosmological constant. Although there is a difference between the evolution in a model with $\Lambda$ and without, these differences are small and by choosing different initial conditions can be minimised.
The model with $\Omega_{mat} = 0.4$ barely reaches the limiting value in the center of the void while models with $\Omega_{mat} \sim 0.3$ fit the observations best.

\section{Conclusions}

The aim of this paper was to build a model of void formation which would
simulate the evolution of voids and would be consistent with observational constraints.
We developed a model that describes the process of void formation from small initial
velocity, density and temperature fluctuations, that existed at the moment of the last scattering, and fully recovers very low values of density contrast inside the voids.
However, in our theoretical model, the present density increases faster at the edges of voids
than in the observed profiles.
There could be several explanations:

\begin{enumerate}
\item
Other shapes of the initial perturbations would reproduce satisfying results,
\item
The assumption (consistent with the widely accepted paradigm) that the distribution of
radiation did not evolve from the last scattering moment is not fulfilled.
\item
 Matter around  voids can  distinctly  depart from spherical symmetry.
\item
The real density contrast increases faster than the density contrast of  luminous matter.
\end{enumerate}

The main conclusion of this paper is that until several million years after  the last scattering moment radiation cannot be neglected  in models of structure formation.
The  gradient of radiation is significant in the process of  void formation.
The negative gradient of radiation causes faster expansion of the space inside the void, hence the density contrast decreases faster there. The excess of radiation pressure simply drives matter out of the region destined to be a void and piles it up on the edges. This effect is  purely relativistic and  in the model presented above radiation does not interact with matter (this is an accurate assumption for the time after the last scattering).
As a result to evolve  structures like voids the amplitude of density fluctuation at the last scattering moment does not have to be larger than $10^{-5}$. Thus, the fluctuation of dark matter at the last scattering can be of the same amplitude as the fluctuation of baryonic matter.

The process of void formation can put some limits on the values of cosmological parameters.
It was found  that models with $\Omega_{mat} \approx 0.3$ describe the present voids best.
In models with larger value of the cosmological constant the density contrast was lower, due to a longer period of evolution. Unfortunately, the only estimation of the density distribution inside the voids is based on the  observations of galaxies, and since in the central parts of voids
no galaxies are observed, there are no precise estimations of density contrast inside them.
Therefore we cannot make any precise conclusions about the value of the cosmological constant.
At present the evolution of voids in both  CDM models, with and without the cosmological constant, is in agreement with the observational data.

\section*{ACKNOWLEDGMENTS}

I would like to thank Andrzej Krasi\'nski for all his help, comments and suggestions while I was preparing the manuscript.  I also thank Paulina Wojciechowska, Fiona Hoyle and Roman Juszkiewicz.

\bsp

\label{lastpage}

\end{document}